\documentclass[11pt,a4paper]{article} 
\usepackage{jheppub}

\usepackage{amsmath, amsthm, amssymb, upref, mathrsfs}

\newcommand{\be}{\begin{equation}}
\newcommand{\ee}{\end{equation}}
\newcommand{\sun}{\odot}

\newcommand{\mpl}{M_\text{Pl}}
\newcommand{\GR}{\text{GR}}
\newcommand{\N}{\text{N}}
\newcommand{\FP}{\text{FP}}
\newcommand{\lbar}{\lambda}

\newcommand{\p}{\partial}
\newcommand{\dd}{\text{d}}

\newcommand{\beq}{\begin{equation}}
\newcommand{\eeq}{\end{equation}}
\newcommand{\bea}{\begin{eqnarray}}
\newcommand{\eea}{\end{eqnarray}}
\newcommand{\sobs}{\sigma_{\rm obs}}

\title{Spherically Symmetric Solutions in Massive Gravity and Constraints from Galaxies} 

\author[a]{Stefan Sj\"ors} 
\author[a]{and Edvard M\"ortsell} 

\affiliation[a]{Department of Physics \& The Oskar Klein Centre for Cosmoparticle Physics, \\
Stockholm University, AlbaNova University Centre,  SE-106 91 Stockholm, Sweden} 

\emailAdd{stefan.sjors@fysik.su.se} 
\emailAdd{edvard@fysik.su.se} 

\abstract{In this paper, analytical solutions describing static and spherically symmetric sources in the decoupling limit of massive gravity are derived. We analyze the model parameter range and specify when a Vainshtein mechanism is possible. Furthermore, we use gravitational lensing and velocity dispersion data from galaxies to put constraints on the mass scale of the graviton. The result for the inverse graviton mass scale $\lambda_g = \hbar /(cm_g)$, in units of the Hubble radius $r_H=c/H_0$, is of the order $\lambda_g/r_H\gtrsim 0.01-0.02$ at 95\,\% confidence level.} 

\keywords{modified gravity, massive gravity, gravitational lensing, velocity dispersion} 

\begin{document} 

\maketitle 

\section{Introduction}

The discovery of the late-time cosmic acceleration \cite{Riess:1998cb, Perlmutter:1998np} has drastically changed our picture of the universe. In the current understanding, the energy content of the universe is dominated by a small cosmological constant, making the cosmological constant problem more pressing than ever. One possible direction towards an explanation of the cosmic acceleration and perhaps also a route to a solution to the cosmological constant problem might be offered by modifications of general relativity (GR) at large distances. As of today, there are many ideas of large distance modifications of GR, some more prolific than others. One simple idea of large distance modifications, one which we pursue in this paper, is the possibility of adding a small mass $m_g$ to the graviton. Needless to say, there are many other interesting scenarios such as the brane world of Dvali-Gabadadze-Porrati (DGP) \cite{Dvali:2000hr} and scalar-tensor theories such as Chameleon \cite{Khoury:2003aq} and $f(R)$ theories \cite{Buchdahl:1983zz}.

These theories have in common that not only do they modify gravity at large distances, but they also introduce small modifications to gravity around local sources. These deviations must necessarily be kept minimal so that the theory passes the very stringent solar system tests. Here, non-linear effects play a crucial role in suppressing the deviations from GR. For example, in massive gravity (MG) and the DGP model, the recovery of GR in regions of strong fields is realized by the so-called Vainshtein mechanism \cite{Vainshtein:1972sx}, while in Chameleon theories, the Chameleon mechanism recovers GR in regions of high density \cite{Khoury:2003aq}. Even though the deviations from GR are kept minimal, all theories give specific predictions for the deviation depending on the non-linear realization of the recovery of GR. This gives us an opportunity to constrain theories of modified gravity using dynamical measurements. 

The deviation from flat space induced by a static spherically symmetric source is in general described by two functions; the gravitational potential $\Phi$ and the spatial curvature $\Psi$. In GR it follows from Einstein's equations that these two functions are the same. Whether they are the same in theories of modified gravity is a question of the specific dynamics. Probing the two potentials using local measurements we can constrain deviations from GR. The two potentials are probed using different methods. Massive non-relativistic observers are sensitive to the gravitational potential $\Phi$, whereas massless relativistic observers are sensitive to the spatial curvature through the combination $\frac{1}{2}(\Phi + \Psi)$.

In this paper we use dynamics of galaxies to constrain deviations from GR. We probe $\Phi$ using velocity dispersion measurements of stars and the combination $\frac{1}{2}(\Phi+\Psi)$ using the gravitational deflection of photons. Gravitational lensing in the context of massive gravity was recently analyzed in \cite{Wyman:2011mp}.

This paper is organized as follows. In \S\ref{sec:background} we give a brief introduction to massive gravity. In \S\ref{sec:formalism} we present the general formalism and the relevant observables, while in \S\ref{sec:model} we discuss the specific model of massive gravity we consider.  In \S\ref{sec:method} we discuss our method and in \S\ref{sec:data} we analyze the galaxy data. We present the results in \S\ref{sec:results} and conclude in \S\ref{sec:conclusions}. We also include an appendix \S\ref{solutions} where we give the details of the solutions to the equations of motion derived in \S\ref{sec:model}.


\section{Background} \label{sec:background}

One of the oldest and simplest ideas of modifications of gravity at large distances is to give the graviton a small mass. The first successful step in this direction was taken by Fierz and Pauli \cite{Fierz:1939ix} already in 1939 when they wrote down a consistent free (linear) theory of massive spin-2 particles. Consistent generalizations to interacting (non-linear) theories turned out to be much harder and examples were always plagued with ghosts. This lead Boulware and Deser to conjecture in 1972 \cite{Boulware:1973my} that there were in fact no consistent non-linear completions of massive gravity. Lately, the subject has spurred renewed interest when a family of actions of non-linear massive gravity was written down in \cite{deRham:2010ik} that was shown to be ghost-free to quartic order in non-linearities, in the so-called decoupling limit \cite{ArkaniHamed:2002sp, Creminelli:2005qk}. Suggestions how to complete the action has since then appeared in the literature \cite{deRham:2010kj} and it was hoped that these actions would be ghost-free to all orders. It was recently shown by Hassan and Rosen \cite{Hassan:2011hr, Hassan:2011ea, Hassan:2011tf} that a certain two-parameter family of non-linear massive gravity actions were indeed ghost-free. Thus it is exciting to consider the phenomenology of massive gravity and specially the explicit implementation of the Vainshtein mechanism.

At the heart of the observational differences between massless and massive gravity lies the van Dam-Veltman-Zakharov (vDVZ) discontinuity \cite{vanDam:1970vg,Zakharov:1970cc}. Na\"ively, one would think that any effects of a graviton mass could only be observed at scales of the inverse mass, where the exponential Yukawa suppression would kick in. In fact, on top of this effect, modifications to the gravitational force law persist even at shorter distances. This lead vDVZ to predict a $3/4$ difference in the angle of light deflected around the sun in MG compared to GR, also in the limit of zero graviton mass. Already at that time, such a big discrepancy rules out FP theory, as the note added in proof reads of \cite{vanDam:1970vg}.

If that was the end of the story, massive gravity would as of then be ruled out observationally. But Vainshtein noticed in \cite{Vainshtein:1972sx} that linear perturbation theory, that is FP theory, breaks down at a distance scale $r_V$, the Vainshtein radius. The Vainshtein radius comprise an intermediate scale between the very small scale of the source, the gravitational radius $r_S \equiv 2 G_\N M/c^2$, and the very large inverse mass scale of the graviton, the reduced Compton wavelength of the graviton $\lbar_g \equiv \hbar /(cm_g)$. Vainshtein showed that outside the Vainshtein radius there exists a perturbation series in $r_V/r$, with FP theory at first order. However, at the Vainshtein radius massive gravity becomes strongly coupled and at smaller scales there exists a small $r/r_V$ limit such that GR is recovered. Whether there existed a non-linear theory that interpolated in between the two regimes was an open question. Numerical work on the subject \cite{Babichev:2009jt} shows that this indeed happens in certain scenarios.

Intuition about the Vainshtein mechanism can be gained by going to the so-called decoupling limit of MG. The decoupling limit offers a calculable regime where we can explicitly interpolate between the regions outside and inside the Vainshtein radius, thus allowing for explicit predictions being made also at the intermediate scale of the Vainshtein radius. The decoupling limit comprise a double-scaling limit taking $\mpl \to \infty$ and $m_g \to 0$, with a certain scale $\Lambda^3 \equiv \mpl \, m_g^2$ fixed. This limit is suitable for our applications since heuristically we can think of this limit as describing massive gravity at distances smaller than the reduced Compton wavelength of the graviton, $\lbar_g \to \infty$, yet on scales large enough that a weak-field limit of general relativity is approximate, $R_S \to 0$. This should be a good approximation since after all gravity is not observed to be exponentially weak at any of the scales we observe today, i.e. we have to push the graviton mass to a very small scale and throughout this paper we work in a regime where all distance scales are much shorter than the reduced Compton wavelength of the graviton, that is $r \ll \lbar_g$. Furthermore, we are always far outside the gravitational radius of any object $r \gg r_S$ where linear gravity is an excellent approximation. 
Numerical solutions to massive gravity in the decoupling limit has been considered in \cite{Chkareuli:2011te} and exact results away from this limit in \cite{Koyama:2011xz, Koyama:2011yg, Nieuwenhuizen:2011sq, Berezhiani:2011mt}.

\section{Formalism} \label{sec:formalism}

The most general static spherically symmetric perturbation of flat space is governed by two functions\footnote{Assuming general covariance.}
\be \label{metricpert}
 \dd s^2 = -(1 + 2 \Phi /c^2) (\dd x^0)^2 + (1 - 2 \Psi /c^2) \dd \mathbf{x}^{2} \,,
\ee
where the gravitational potential $\Phi$ and spatial curvature $\Psi$ are functions of the radial distance $r = | \mathbf{x} |$ alone, and as usual we put $x^{\mu} = (ct, \mathbf{x})$, $\mu=0,1,2,3$. In linearized GR the two potentials coincide $\Phi = \Psi$, but in theories of modified gravity the potentials are no longer equal.
Within the parametrized post-Newtonian (PPN) formalism, the deviation from Newtonian gravity is quantified in terms of the ratio of the two potentials $\Psi/\Phi$. For a constant ratio, this defines the parameter $\gamma_{\rm PPN} = \Psi/\Phi$, with $\gamma_{\rm PPN} = 1$ in linearized GR. In general, deviations from GR are not captured by a single number but rather two functions. 

The two potentials can be probed using different methods. A massive non-relativistic observer is only sensitive to the gravitational potential $\Phi$ while a massless relativistic observer is sensitive also to the curvature $\Psi$. Specifically, a massless observer is sensitive to the potential combination $\Phi_+ \equiv \frac{1}{2} (\Phi + \Psi)$, known as the lensing potential. Thus complementary observations using massive and massless probes can access both potentials and allows us to quantify deviations from GR. 

More quantitatively, we analyze the above geometry, eq. (\ref{metricpert}), using a probe with energy-momentum tensor $\tau_{\mu\nu}$. For a point particle moving on a trajectory $\mathbf{x} = \mathbf{x}(t)$, the energy-momentum tensor is given by
\be
 \tau^{\mu\nu} (\mathbf{x}') = c \, \frac{p^\mu p^\nu}{p^0} \delta^{(3)}(\mathbf{x}' - \mathbf{x}(t)) \,.
\ee
The force felt by the probe is $\mathbf{F}(\mathbf{x}) = - \nabla U(\mathbf{x})$ where $U(\mathbf{x})$ is the interaction energy between the probe and the background. For a probe that is minimally coupled to gravity the interaction energy is given by
\be \label{interaction_energy}
 U(\mathbf{x}) = -\frac{1}{2} \int \dd^3 \mathbf{x}' \, h^{\mu\nu}(\mathbf{x}') \, \tau_{\mu\nu}(\mathbf{x}') \,,
\ee
where $h_{\mu\nu} = g_{\mu\nu} - \eta_{\mu\nu}$ is the deviation of the metric, in eq. (\ref{metricpert}), from flat space. For a non-relativistic, nearly stationary particle of mass $m$ and four-momentum $p^{\mu} = (mc; \mathbf{0})$ we pick up $h_{00} = - 2\Phi/c^2$ in the above and get for the interaction energy $U = - m \Phi$. This reproduces the standard force law of a non-relativistic observer
\be
 \mathbf{F} = - m \nabla \Phi \,.
\ee
For a massless particle $p^2 = 0$ the energy momentum tensor is traceless $\tau_{\mu}^{\phantom{\mu}\mu} = 0$ and we are only sensitive to the traceless part of the metric perturbation which contains the combination $\Phi_+ = \frac{1}{2}(\Phi+\Psi)$, indeed $h^\text{trace-free}_{\mu\nu} = h_{\mu\nu} - \frac{1}{4} \eta_{\mu\nu} h = -\Phi_+ / c^2 \cdot \text{diag} (3,1,1,1)$.
Taking for example $p^\mu = (E/c; 0,0,E/c)$ the interaction energy becomes $U = - 2E\Phi_+ /c^2$.

\subsection{Massless and massive linear gravity}

As an example, consider a point source with mass $M$. In the weak-field limit of GR, the two potentials in eq. (\ref{metricpert}) are the same and equals the Newtonian potential $\Phi_\N \equiv - G_\N M/r$ of the point source
\be
 \Phi|_\GR = \Psi|_\GR = - \frac{G_\N M}{r} \,.
\ee
Thus, in GR $\gamma = \Psi/\Phi = 1$. As noticed by vDVZ, the situation is quite different in FP theory of massive gravity where\footnote{Up to gauge equivalent terms.}
\be \label{linearsolution}
 \Phi|_\FP = - \frac{4G_\N M}{3r} e^{-r/\lbar_g} \,, \quad \Psi|_\FP = - \frac{2 G_\N M}{3r} e^{-r/\lbar_g} \,,
\ee
see for example the nice review \cite{Hinterbichler:2011tt}. At long distances, gravity is exponentially weak being cut-off by the graviton mass whereas at shorter distances, the gravitational force is effectively $1/r^2$. A massive observer probing short distances therefore experiences a Newtonian potential $\Phi|_\FP = - G'_\N M/r$, albeit with a modified Newton's constant $G'_\N=4G_\N/3$. The degeneracy with GR is lifted using massless observers who are sensitive also to the curvature perturbation $\Psi$ and Fierz-Pauli theory predicts $\gamma|_\FP = (\Psi/\Phi)|_\FP = 1/2$. More strikingly put, since the lensing potential is the same in massive gravity as in massless, $\Phi_+|_\FP = \frac{1}{2}(\Phi+\Psi)|_\FP = \Phi_\N$, the prediction in terms of the modified Newton's constant of the deflection angle of a light-ray passing the perimeter of the sun is a factor of $3/4$ different in massive gravity than in massless, regardless of the graviton mass
\be
 \theta_\sun = \frac{4G_\N M_\sun}{r_\sun} = \frac{3}{4} \times \frac{4G'_\N M_\sun}{r_\sun} \,.
\ee

We choose to quantify the deviations from the force laws in GR defining two functions $\varepsilon_\Phi(r),\varepsilon_\Psi(r)$, such that $\varepsilon_\Phi, \varepsilon_\Psi = 0$ defines GR,
\begin{align}
 \nabla \Phi(r) & \equiv [1 + \varepsilon_\Phi(r)] \, \nabla \Phi_\N(r) \,, \label{epsilon} \\
 \nabla \Psi(r) & \equiv [1 - \varepsilon_\Psi(r)] \, \nabla \Phi_\N(r) \,. \label{eta}
\end{align}
Notice that $\varepsilon_{\Phi}, \varepsilon_{\Psi}$ effectively changes the Newton's constant we would infer in a local measurement. For example, a non-relativistic observer would effectively experience a slightly varying Newton's constant $G_\N^\mathrm{m}(r) = [1+\varepsilon_\Phi(r)]G_\N$ while a relativistic observer would see a Newton's constant $G_\N^\gamma(r) = (1+[\varepsilon_\Phi(r) - \varepsilon_\Psi(r)]/2) G_\N$.

Notice that in a phenomenologically viable theory of modified gravity, $\varepsilon_\Phi(r)$ and $\varepsilon_\Psi(r)$ necessarily vanish at short distances such that solar system constraints are evaded. For example, within the decoupling limit of massive gravity, $\varepsilon_\Phi(r)$ and $\varepsilon_\Psi(r)$ are slowly varying functions that vary from zero deep inside the Vainshtein radius, reaching one third at scales far outside the Vainshtein radius, i.e.
\be \label{boundaryconditions}
 \lim_{r/r_V \to 0} \varepsilon_\Phi(r),\varepsilon_\Psi(r) = 0 \,, \quad
 \lim_{r/r_V \to \infty} \varepsilon_\Phi(r), \varepsilon_\Psi(r) = 1/3 \,.
\ee

\section{Model} \label{sec:model}
An intermediate step towards a ghost-free theory of massive gravity was taken in \cite{deRham:2010ik}. Here massive gravity was analyzed in the so-called decoupling limit and a certain two-parameter family of actions were shown to be ghost-free around flat space. The decoupling limit comprise a calculable approximation of massive gravity and corresponds to taking the limit $\mpl \to \infty$ and $m_g \to 0$, with a certain scale $\Lambda^3 \equiv \mpl \, m_g^2$ fixed. This limit is suitable for our applications since heuristically we can think of this limit as describing massive gravity at distances smaller than the reduced Compton wavelength of the graviton, $\lbar_g \to \infty$, yet on scales large enough that a weak-field limit of general relativity is approximate, $r_S \to 0$.

\subsection{Action and equations of motion}

We now discuss the action of massive gravity in the decoupling limit. At the linearized level, the theory describes a massive spin-2 field with five degrees of freedom. In the decoupling limit, where the mass is taken to zero, the two helicity-1 modes become longitudinal (in the four-vector sense) and decouple from any conserved source. What remains is the action of the two helicity-2 modes and the one helicity-0 mode. Following \cite{deRham:2010tw} the action is given by\footnote{Throughout \S\ref{sec:model}, and only this section, we use natural units and put $\hbar = c = 1$.}
\be \label{action}
 \mathscr{L} = - \frac{1}{2} H^{\mu\nu} \mathcal{E}_{\mu\nu}^{\alpha\beta} H_{\alpha\beta} + H^{\mu\nu} \left( \alpha X^{(1)}_{\mu\nu} + \frac{\beta}{\Lambda^3} X^{(2)}_{\mu\nu} + \frac{\gamma}{\Lambda^6} X^{(3)}_{\mu\nu} \right) + \frac{1}{2\mpl} H^{\mu\nu} T_{\mu\nu} \,.
\ee
The first term in eq. (\ref{action}) comes from evaluating the Einstein-Hilbert action $\mpl^2 \sqrt{-g} R$ to quadratic order in the metric fluctuation $\mpl^{-1} H_{\mu\nu} = g_{\mu\nu} - \eta_{\mu\nu}$. Explicitly, the so-called Einstein operator $\mathcal{E}_{\mu\nu}^{\alpha\beta}$ takes the form

\be
 \mathcal{E}_{\mu\nu}^{\alpha\beta} H_{\alpha\beta} = -\frac{1}{2}\left( \p^2 H_{\mu\nu} - 2 \p_{(\mu} \p^\rho H_{\nu)\rho} + \p_\mu \p_\nu H - \eta_{\mu\nu} \p^2 H + \eta_{\mu\nu} \p_\alpha \p_\beta H^{\alpha\beta} \right).
\ee
The second term in eq. (\ref{action}) is a linear coupling of $H_{\mu\nu}$ and $\pi$ consistent with the Galilean symmetry $\pi \to \pi + c + b \cdot x$, forcing $\pi$ to enter in the combination $\Pi_{\mu\nu} \equiv \p_\mu \p_\nu \pi$, together with gauge invariance $H_{\mu\nu} \to H_{\mu\nu} - 2 \p_{(\mu} \xi_{\nu)}$, forcing $X_{\mu\nu}^{(1,2,3)}$ to be divergence free. Explicitly
\begin{align}
 X^{(1)}_{\mu\nu} & = \varepsilon_{\mu}^{\phantom{\mu}\alpha\rho\sigma} \varepsilon^{\phantom{\nu}\beta}_{\nu\phantom{\beta}\rho\sigma} \Pi_{\alpha\beta} \,, \\
 X^{(2)}_{\mu\nu} & = \varepsilon_{\mu}^{\phantom{\mu}\alpha\rho\sigma} \varepsilon_{\nu\phantom{\beta\gamma}\sigma}^{\phantom{\nu}\beta\gamma} \Pi_{\alpha\beta} \Pi_{\rho\gamma} \,, \\
 X^{(3)}_{\mu\nu} & = \varepsilon_{\mu}^{\phantom{\mu}\alpha\rho\sigma} \varepsilon_{\nu}^{\phantom{\nu}\beta\gamma\delta} \Pi_{\alpha\beta} \Pi_{\rho\gamma} \Pi_{\sigma\delta} \,.
\end{align}
The term $\alpha \, H^{\mu\nu} X^{(1)}_{\mu\nu}$ gives a kinetic mixing between $H_{\mu\nu}$ and $\pi$, and positivity of the kinetic energy gives $\alpha  < 0$. Apart from the sign, $\alpha$ is just a normalization of the field $\pi$. The parameters $\beta,\gamma$ are free model parameters. Setting $\alpha, \beta, \gamma = 0$ recovers weak-field GR. The third term in eq. (\ref{action}) is the interaction energy density of source and gravity, as discussed in connection to eq. (\ref{interaction_energy}).

From the above action, eq. (\ref{action}), we can derive the metric equations of motion
\be \label{metriceom}
 \mathcal{E}_{\mu\nu}^{\alpha\beta} H_{\alpha\beta}  = \alpha X^{(1)}_{\mu\nu} + \frac{\beta}{\Lambda^3} X^{(2)}_{\mu\nu} + \frac{\gamma}{\Lambda^6} X^{(3)}_{\mu\nu} + \frac{1}{2\mpl} T_{\mu\nu} \,,
\ee
and the $\pi$ equation of motion
\be \label{pieom1}
 \p_\alpha \p_\beta H^{\mu\nu} \left(
 \alpha \, \varepsilon_{\mu}^{\phantom{\mu}\alpha\rho\sigma} \varepsilon^{\phantom{\nu}\beta}_{\nu\phantom{\beta}\rho\sigma}
 + \frac{2\beta}{\Lambda^3} \varepsilon_{\mu}^{\phantom{\mu}\alpha\rho\sigma} \varepsilon_{\nu\phantom{\beta\gamma}\sigma}^{\phantom{\nu}\beta\gamma} \Pi_{\rho\gamma}
 + \frac{3 \gamma}{\Lambda^6} \varepsilon_{\mu}^{\phantom{\mu}\alpha\rho\sigma} \varepsilon_{\nu}^{\phantom{\nu}\beta\gamma\delta} \Pi_{\rho\gamma} \Pi_{\sigma\delta} \right) = 0 \,.
\ee

\subsection{Static and spherically symmetric ansatz}

In this section we analyze the equations for a static, spherically symmetric, and pressure-free source described by a mass density $\rho=\rho(r)$. The most general ansatz consistent with these symmetries can be put on the form
\begin{align}
 \dd s^2 & = -[1 + 2 \Phi(r)] \, \dd t^2 + [1 - 2 \Psi(r)] \, \dd \mathbf{x}^2 \,, \label{metric} \\
 \pi & = \pi(r) \,, \label{pi} \\
 T_{\mu\nu} & = \text{diag}(\rho(r),0,0,0) \,. \label{stress}
\end{align}
We now substitute the above ansatz into the equations of motion. Of the ten equations of motion of the metric, only the time-time and radial-radial component equations give independent equations. Together with the $\pi$ equation of motion we have three independent equations for three unknowns and the system should admit a solution.

The two metric equations of motion read (using $'= \frac{\dd}{\dd r}$ to denote a radial derivative)
\begin{align}
 \mpl \, \frac{2}{r^2} (r^2 \Psi')' & = \frac{\rho}{2 \mpl} + \alpha \, \frac{2}{r^2} (r^2 \pi')' + \frac{\beta}{\Lambda^3} \, \frac{2}{r^2} (r (\pi')^2)' + \frac{\gamma}{\Lambda^6} \, \frac{2}{r^2} ((\pi')^3)' \,, \label{timeeom} \\
 \mpl \, \frac{2}{r} (\Phi' - \Psi') & = - \alpha \, \frac{4}{r} \pi' - \frac{\beta}{\Lambda^3} \frac{2}{r^2} (\pi')^2 \,,\label{radialeom}
\end{align}
while the $\pi$ equation of motion reads
\be
  \left[ \alpha r^2 (2 \Psi' - \Phi') + \frac{2 \beta}{\Lambda^3} r \pi' (\Psi' - \Phi') - \frac{3 \gamma}{\Lambda^6} \Phi' (\pi')^2 \right]' = 0 \,. \label{pieom}
\ee

First, note that eq. (\ref{timeeom}) is a total derivative and can be integrated with the result
\be \label{gauss}
 \Psi' = \frac{1}{16 \pi \mpl^2} \frac{M(r)}{r^2} + \frac{\alpha}{\mpl} \pi' + \frac{\beta}{\mpl \Lambda^3} \frac{(\pi')^2}{r} + \frac{\gamma}{\mpl\Lambda^6} \frac{(\pi')^3}{r^2} \,,
\ee
where $M(r)$ denotes the integrated mass inside the sphere of radius $r$
\be \label{mass}
 M(r) = 4 \pi \int_0^r \dd R \, R^2 \rho(R) \,.
\ee
This result is a manifestation of `Gauss' law' or the `Shell theorem': The force experienced at a radius $r$ only depend on the integrated mass inside that radius, and the force is equivalent to that of a point particle of mass $M$ situated at the origin. This property is an artifact of decoupling limit of MG, where the exponential Yukawa decay is pushed to infinity. Certainly, this is only a property of $1/r^2$ forces and in the full theory this result does not hold.

We now discuss briefly the two sub-cases GR: $\alpha,\beta,\gamma=0$ and FP: $\alpha < 0,\,\beta,\gamma=0$ before going on to the general case.

\paragraph{GR case:}

From eq. (\ref{radialeom}) we see that in the GR limit, putting $\alpha,\beta,\gamma = 0$, then indeed $\Phi' - \Psi' = 0$. Using eq. (\ref{gauss}) we reproduce the standard Newtonian force law
\be
 \Phi'|_\GR = \Psi'|_\GR = \frac{1}{16 \pi \mpl} \, \frac{M(r)}{r^2} \equiv \Phi'_\N \,,
\ee
where we in the last step identified the Newtonian force $\Phi'_\N \equiv G_\N M(r) /r^2$, with Newton's constant\footnote{Note our normalization of $\mpl$ which is different from other normalizations often used in the literature.} $G_\N \equiv (16 \pi \mpl)^{-1}$.

\paragraph{FP case:}

In the case $\alpha < 0$, $\beta,\gamma=0$, integrating eq. (\ref{pieom}) forces $2 \Psi' - \Phi' = 0$, (setting the integration constant to zero). Then using eqs. (\ref{radialeom}) and (\ref{gauss}) we find
\be
 \Phi'|_\FP = \frac{3}{4} \Phi'_\N \,, \quad \Psi'|_\FP = \frac{2}{3} \Phi'_\N \,, \quad -\frac{\alpha}{\mpl} \pi'|_\FP = \frac{1}{3} \Phi'_\N \,,
\ee
which reproduces eq. (\ref{linearsolution}), (which also verifies the correct choice of integration constant).

\paragraph{General case:}

Away from the GR limit we parametrize the deviations in the force laws using the two functions, $\varepsilon_\Phi = \Phi'/\Phi'_\N - 1$ and $\varepsilon_\Psi = 1 - \Psi' / \Phi'_\N$, defined in eqs. (\ref{epsilon}) and (\ref{eta}). Using eqs. (\ref{radialeom}) and (\ref{gauss}) we find
\begin{align}
\varepsilon_\Phi & = -\frac{\alpha}{\mpl} \frac{\pi'}{\Phi_N'} + \frac{\gamma}{\mpl\Lambda^6} \frac{(\pi')^3}{r^2\Phi_\N'} \,, \label{epsilonphi} \\
\varepsilon_\Psi & = - \frac{\alpha}{\mpl} \frac{\pi'}{\Phi_N'} - \frac{\beta}{\mpl\Lambda^3}\frac{(\pi')^2}{r \Phi_\N'} - \frac{\gamma}{\mpl\Lambda^6} \frac{(\pi')^3}{r^2 \Phi_\N'} \,. \label{epsilonpsi}
 \end{align}
Thus, the deviations are completely determined by the behavior of the $\pi$ field, or more specifically by powers of $\pi'$. Substituting the expressions for $\Psi'$ and $\Phi'$ into eq. (\ref{pieom}) for $\pi$, we find a closed algebraic equation for $\pi'$. Solving this equation for $\pi'$ then determines $\Phi'$ and $\Psi'$ completely. Indeed, integrating eq. (\ref{pieom}) (setting the integration constant to zero) and using the expression for $\Psi'$ and $\Phi'$ we find a closed algebraic equation for $\pi'$
\begin{align} \label{quintic}
 \alpha r^2 \Phi'_\N + \frac{3 \alpha^2 r^2}{\mpl} \pi' & + \left( \frac{6\alpha\beta r}{\Lambda^3 \mpl} - \frac{3\gamma \Phi'_\N}{\Lambda^6} \right) (\pi')^2 \nonumber \\
  & + \frac{2\beta^2 + 4\alpha\gamma}{\Lambda^6 \mpl} (\pi')^3 - \frac{3\gamma^2}{\Lambda^{12} \mpl r^2}(\pi')^5 = 0 \,.
\end{align}

\subsection{Solutions for $\pi'$}
We now discuss the general features of the solutions to eq. (\ref{quintic}), relegating the details to \S\ref{solutions} in the appendix. First, note that eq. (\ref{quintic}) is a quintic polynomial equation in $\pi'$ and will therefore in general not allow for closed-form solutions\footnote{Note that as a polynomial in $r$, eq. (\ref{quintic}) is actually a quartic and $r=r(\pi')$ can be obtained in a closed form.}. However, as we show in \S\ref{solutions} the quintic term is negligible in almost all of the parameter space $[\beta,\gamma]$ and we can drop the $(\pi')^5$ term from eq. (\ref{quintic}), thus obtaining a cubic equation which we solve analytically.

\begin{figure}[htbp]
\begin{center}
\includegraphics[width=0.496\textwidth]{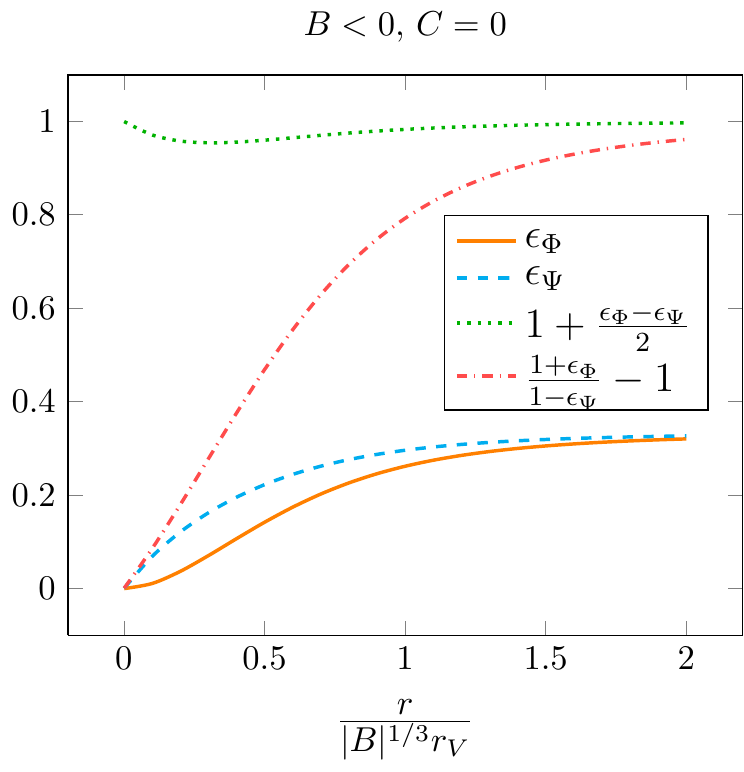}
\includegraphics[width=0.496\textwidth]{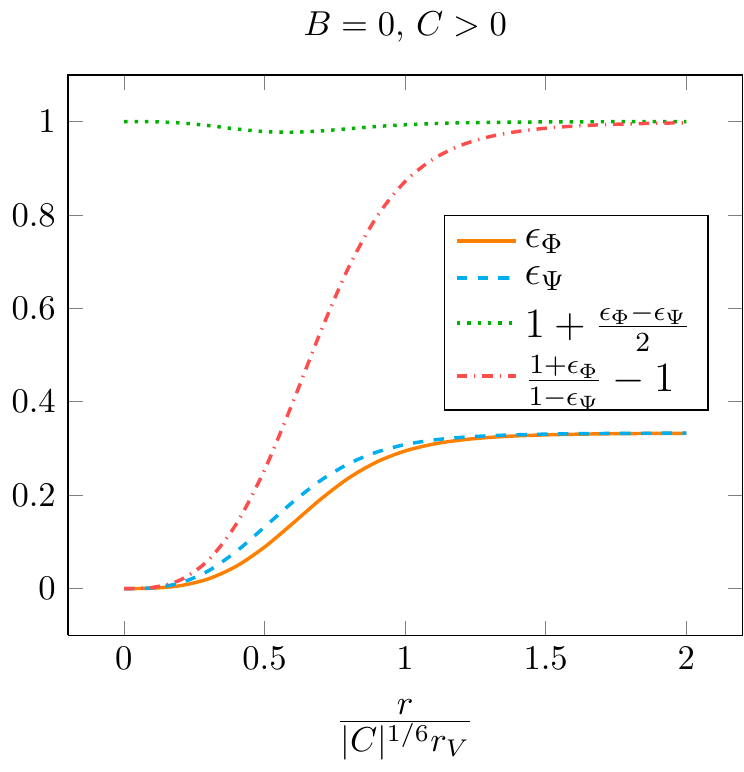}
\caption{Left: Results for $C=0$ and $B<0$. Right: Results for $B=0$ and $C>0$. We plot the deviations in the force laws $\varepsilon_\Phi = \Phi'/\Phi'_\N - 1$ and $\varepsilon_\Psi = 1 - \Psi'/\Phi'_\N$, together with deviation in the lensing potential $\Phi'_+/\Phi'_\N = 1 + \frac{\varepsilon_\Phi - \varepsilon_\Psi}{2}$ and the ratio $\Phi'/\Psi' - 1 = \frac{1+\varepsilon_\Phi}{1-\varepsilon_\Psi}-1$. The solutions exhibit the Vainshtein mechanism, eq. (\ref{boundaryconditions}), i.e. $\lim_{r/r_V \to 0} \varepsilon_\Phi(r),\varepsilon_\Psi(r) = 0$, $\lim_{r/r_V \to \infty} \varepsilon_\Phi(r), \varepsilon_\Psi(r) = 1/3$. Furthermore, the lensing potential is essentially that of GR everywhere with small corrections at $r \simeq r_V$, i.e. $1 + \frac{\varepsilon_\Phi - \varepsilon_\Psi}{2} \simeq 1$ almost everywhere.}
\label{plotsolution}
\end{center}
\end{figure}

For $\gamma = 0$, we find two solutions depending on the sign of
\be 
 B \equiv \beta/\alpha^2 \,.
\ee
For $\text{sign}(B) = +1$, the solution blows up at infinity and does not satisfy the right boundary conditions. For $\text{sign}(B) = -1$, we find a solution with all the right boundary conditions as specified in eq. (\ref{boundaryconditions}), allowing for a Vainshtein mechanism. The analytic solution for $\gamma=0$ and with $\text{sign}(B)=-1$ is presented in eq. (\ref{betasolution}) and plotted in Figure \ref{plotsolution}. The figure clearly demonstrates the Vainshtein mechanism; interpolating between GR at short distances $r \ll r_V$ and FP at long distances $r \gg r_V$.

For $\beta = 0$ we likewise find two solutions depending this time on the sign of
\be
 C \equiv \gamma/\alpha^3 \,.
\ee
For $\text{sign}(C) = -1$, the solution blows up while for $\text{sign}(C) = +1$, there is a solution with all the right boundary conditions, as specified in eq. (\ref{boundaryconditions}). The analytic solution for $\beta=0$ with $\text{sign}(C)=+1$ is presented in eq. (\ref{gammasolution}) and plotted in Figure \ref{plotsolution}.

The solution for non-zero $\beta,\gamma$ is presented in eqs. (\ref{generalsolution}) and (\ref{generalsolution2}). Eq. (\ref{generalsolution}) covers the range $C>0$, $B<0$ while eq. (\ref{generalsolution2}) covers the range $C>0$, $0 < B \lesssim B_\mathrm{max}$. At $B_\mathrm{max} \equiv \sqrt{5-\sqrt{13}}\sqrt{2C}$ the quintic term gives important contributions and a singular behavior is developed beyond $B_\mathrm{max}$, see Figure \ref{fig:bound} in \S\ref{solutions}. We conclude that the quintic equation (\ref{quintic}) allows for solutions with a Vainshtein mechanism, satisfying eq. (\ref{boundaryconditions}), in the parameter range $C\geq0$ with $B < B_\mathrm{max}$. Furthermore, the solutions to the cubic equation, neglecting the quintic term, approximates the solutions to the quintic equation within $\mathcal{O}(10^{-6})$ in all of the parameter space $C>0$, $B \lesssim B_\mathrm{max}$, except at $B \simeq B_\mathrm{max}$ where the quintic becomes important, and there are up to $40\%$ corrections to the cubic solution.


\section{Method} \label{sec:method}

We now turn to the observational consequences of the above considered models. As explained in \S\ref{sec:formalism} we use the fact that a non-relativistic and a relativistic observer experience different effective gravitational couplings, $G_\N^\mathrm{m}(r) = [1+\epsilon_\Phi(r)] G_\N$ and $G_\N^\gamma(r) = (1+ \frac{1}{2}[\epsilon_\Phi(r)-\epsilon_\Psi(r)]) G_\N$ respectively.
Thus, deviations from GR can be probed by comparing the masses of galaxies as measured from gravitational lensing and the velocity dispersion of stars within the galaxies. In practice, we compare the observed velocity dispersion within galaxies with the theoretically expected velocity dispersion assuming a model for the luminosity and mass distribution where the normalization of the mass is set by the image separation of the lensed images. The theoretically inferred velocity dispersion will thus include the modified gravitational dynamics for both non-relativistic and relativistic observers.

In the following, we will to a large extent follow the methods
outlined in \citet{2010ApJ...708..750S} (see also
\citet{2010PhRvD..81j3002S}). We will assume that the mass densities $\rho$ and
the luminosity densities $\nu$ of the lensing galaxies can be written as power laws
\beq
  \rho=\rho_0r^{-\alpha}\,, \quad \nu=\nu_0r^{-\delta}\,.
\eeq
The spherical mass inside radius $r$ is then
\beq
  M(r)= 4 \pi \int_0^r \dd R \, R^2 \rho(R) = \frac{4\pi\rho_0}{3-\alpha}r^{3-\alpha}\, ,
\eeq
and the projected cylindrical mass traced by lensing within radius $R$
\beq
  M_\text{proj}(R)=\frac{2\pi^{3/2}\lambda(\alpha)\rho_0}{3-\alpha}R^{3-\alpha}\, ,
\eeq
where $\lambda(x)$ is the ratio of gamma functions $\lambda(x) = \Gamma (\frac{x-1}{2}) / \Gamma(\frac{x}{2})$. Specifically, the projected mass within the Einstein radius $R_E$ is 
\beq
  M_\text{proj}(R_E)=\frac{2\pi^{3/2}\lambda(\alpha)\rho_0}{3-\alpha}R_E^{3-\alpha}\, .
\eeq
Since the angular separation of lensing images provides a good
approximation to the corresponding Einstein angle $\theta_E=R_E/D_{l}$ we can
use lensing data to fix the normalization of the galaxy mass using
\beq
  M_\text{proj}(R_E)=\frac{c^2}{4G_\N^\gamma(R_E)}\frac{D_{s}}{D_{ls}D_{l}}R_E^2\, ,
  \label{eq:RE}
\eeq
where $D_{l}, D_{s}$ and $D_{ls}$ are angular diameter distances
between the observer and lens, the observer and source, and the lens
and source, respectively.  Note in eq.~(\ref{eq:RE}), we use the appropriate effective Newton's constant $G^\gamma_\N(r)$ being relevant for light. Furthermore, when deriving eq. (\ref{eq:RE}) we have used that $G_\N^\gamma(r)$ is slowly varying such that when evaluating the light deflection line-of-sight integral, $G^\gamma_\N(r)$ can be approximated by $G^\gamma_\N(R_E)$.

We note that the ratio of the Einstein radius $R_E$ and the Vainshtein
radius $r_V$ is given by
\beq
  \rho_E^3=\left(\frac{R_E}{r_V}\right)^3=\frac{D_{ls}D_{l}^2}{D_{s}\lambda_g^2}\pi\theta_E\, ,
\eeq
which is $\ll1$ unless $\lambda_g \ll D_i$.

\subsection{Velocity dispersion}
Equations of stellar hydrodynamics give for the radial velocity dispersion
\beq
  \sigma_r^2(r)=\frac{1}{r^{2\beta}\nu(r)}\int_r^\infty G^\mathrm{m}_\N(r)\nu(r)M(r)r^{2\beta-2}\, ,
\eeq
where $\beta = 1-(\sigma_t/\sigma_r)^2$ is the (constant) velocity anisotropy of the system\footnote{Note that following conventions, we use the same notations $\alpha$ and $\beta$ in describing the density profile and the velocity anisotropy as used as coefficients in the action eq. (\ref{action}).} and $G^\mathrm{m}_\N(r)$ is the effective gravitational coupling as felt by massive observers. Since $G^\mathrm{m}_\N(r) = [1 + \epsilon_\Phi(r)] G_\N$, we can write the velocity dispersion as a sum of the familiar Newtonian expression and a term depending on deviation in the force law $\epsilon_\Phi(r)$
\beq
  \sigma_r^2(r)=\sigma_\N^2(r)+\sigma^2_{\epsilon_\Phi}(r)\, ,
\eeq
where $\sigma_\N^2(r)$ is the standard Newtonian result and $\sigma^2_{\epsilon_\Phi}(r)$ is given by
\beq
  \sigma^2_{\epsilon_\Phi}(r)=\frac{G_\N}{r^{2\beta}\nu(r)}\int_r^\infty \epsilon_\Phi(r) \nu(r) M(r)r^{2\beta-2}\, .
\eeq

The observed velocity dispersion, $\sigma_*$, is effectively
luminosity-weighted along the line of sight and over the effective
spectrometer aperture. This averaging can be expressed as
\begin{equation}\label{eq:sobs}
  \sigma^2_*=\frac{\int_0^{R_{\rm max}} dR \ R \ w(R)
  \int_{-\infty}^{\infty} d\mathcal{Z} \ \nu(r) \ \left(1 - \beta
  \frac{R^2}{r^2}\right)\sigma^2_r(r)}{\int_0^{R_{\rm max}} dR \ R \ w(R)
  \int_{-\infty}^{\infty} d\mathcal{Z} \ \nu(r)}\, ,
\end{equation}
where $\mathcal{Z}^2=r^2-R^2$ and
\begin{equation}
  w(R) \approx  e^{-R^2/2 \tilde{\sigma}_{\rm atm}^2}\,  ,
\label{eq:wofr}
\end{equation}
is the aperture weighting function in which we will use a median
seeing of $\tilde{\sigma}_{\rm atm}=1.4''$ and a cut-off radius of
$R_{\rm max}=3''$ \cite{2008ApJ...682..964B}.  In eq.~(\ref{eq:sobs}),
the factor $\left(1-\beta \frac{R^2}{r^2}\right)$ takes into account
how the radial and tangential components of the velocity dispersion
project along the line of sight.

\section{Data} \label{sec:data}
We use the data for the full sample of 131 strong gravitational lens
candidates observed with the Advanced Camera for Surveys (ACS) aboard
the Hubble Space Telescope (HST) by the Sloan Lens ACS (SLACS) Survey
\cite{2008ApJ...682..964B}. The lensing foreground galaxies are
primarily of early-type morphology, with redshifts $0.05<z<0.5$ and
velocity dispersions $160\,\mathrm{km/s}<\sobs<400 \mathrm{km/s}$. The lensed background
galaxies have redshifts $0.2<z<1.2$. The SLACS lens
sample is statistically consistent with being drawn at random from a
parent sample of comparable Sloan Digital Sky Survey (SDSS) galaxies,
although our analysis does not depend on this property. Out of these
systems, we use a subsample of 53 systems with elliptical lensing
galaxies where singular isothermal ellipsoid gravitational lens models 
can successfully be fitted to the imaging
data and we have reliable velocity dispersion measurements.

For the present work, the key observables in each system are the
redshifts and the stellar velocity dispersion of the lens galaxy (as
measured from SDSS spectroscopy) and the Einstein angle of
the strongly lensed image of the background galaxy (as measured from
ACS imaging). We also make use of the effective radius measured from
ACS imaging data to obtain individual estimates of the luminosity
profile of the lensing galaxy. Following \cite{2008ApJ...682..964B},
we adopt a 7\,\% velocity dispersion uncertainty and a 2\,\% Einstein
radius uncertainty. We also add an additional 5\,\% to the velocity
uncertainty given that we do not expect the observed velocity dispersion to
perfectly match the one expected for a Singular Isothermal mass profile
\cite{2008A&A...477..397G}. Note that the choice of luminosity profile
may potentially alter the results, e.g. going from a power law to a
Hernquist profile
\cite{2009arXiv0907.4829S}.


\section{Results} \label{sec:results}
We are now in a position to generate constraints on the massive gravity model by comparing the
observed velocity dispersions $\sobs$
with the theoretically calculated velocity dispersions for the given galaxy 
and massive gravity model
$\sigma_*$. From the analysis in the previous section, we have
\begin{equation}
  \sigma_* = \sigma_* (\alpha, \beta, \delta, \theta_E; \lambda_g, B, C)\, ,
\label{eq:sigma_pars}
\end{equation}
where the semicolon separates the galaxy parameters and observables
($\alpha$, $\beta$, $\delta$, $\theta_E$) from the global massive gravity parameters
($\lambda_g, B,C$) which we are seeking to constrain.
The slope and anisotropy of the lensing galaxies, $\alpha$ and $\beta$, are
being marginalized over on an individual basis for each galaxy
assuming a prior knowledge of $\alpha=2.00\pm 0.08$ and $\beta=0.13\pm
0.13$ (68\,\% confidence level) \cite{2010ApJ...708..750S}. We also
make use of the effective radii measured from ACS imaging data to
obtain individual estimates of the luminosity profile power-law index, $\delta$, 
by comparing the observed total luminosity to the luminosity within
half the effective  radius.

\begin{figure}
\begin{center}
\includegraphics[angle=0,width=.49\textwidth]{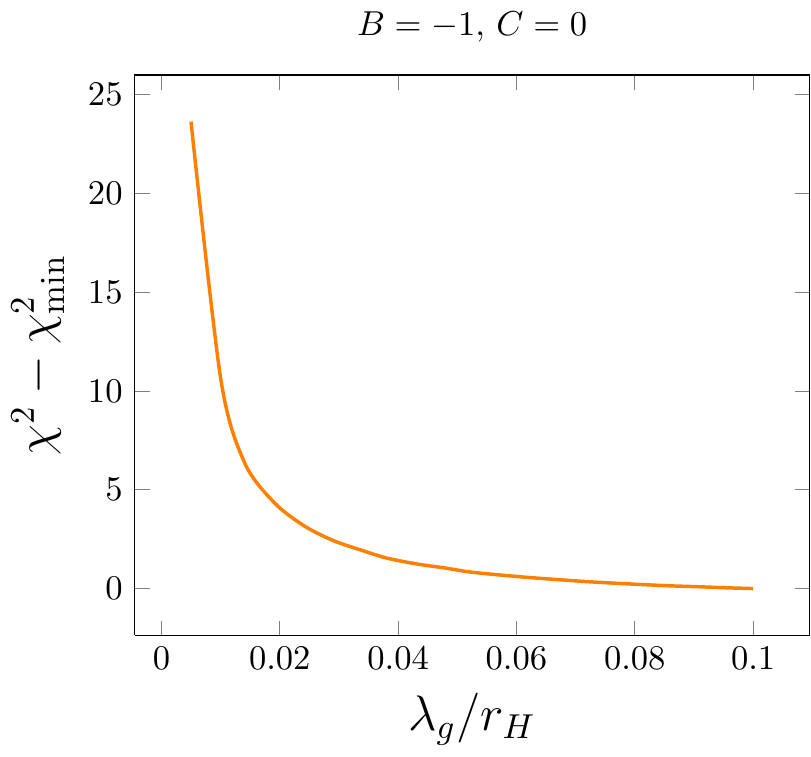}
\includegraphics[angle=0,width=.49\textwidth]{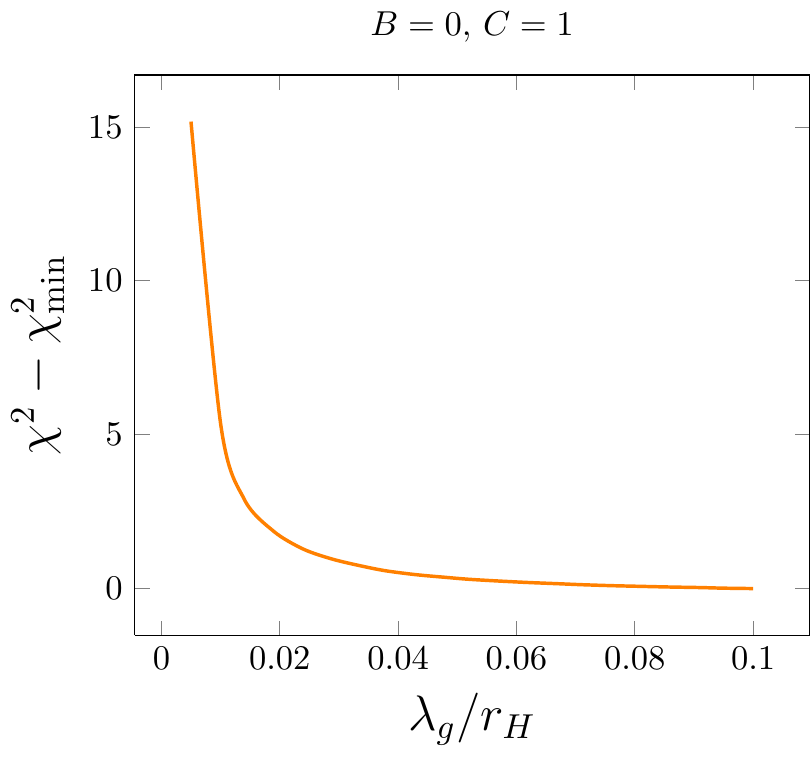}
\includegraphics[angle=0,width=.49\textwidth]{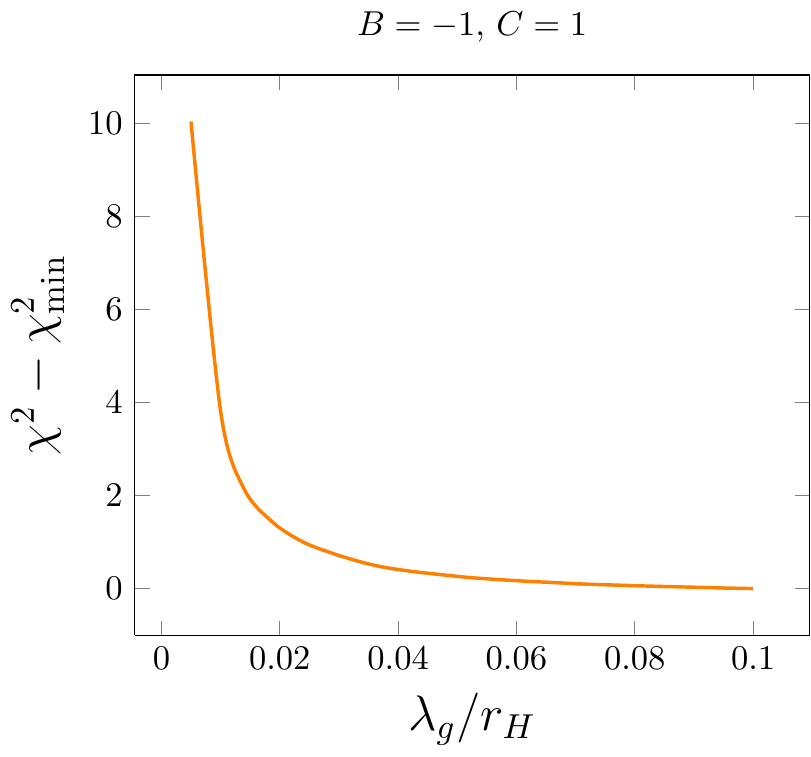}
\includegraphics[angle=0,width=.49\textwidth]{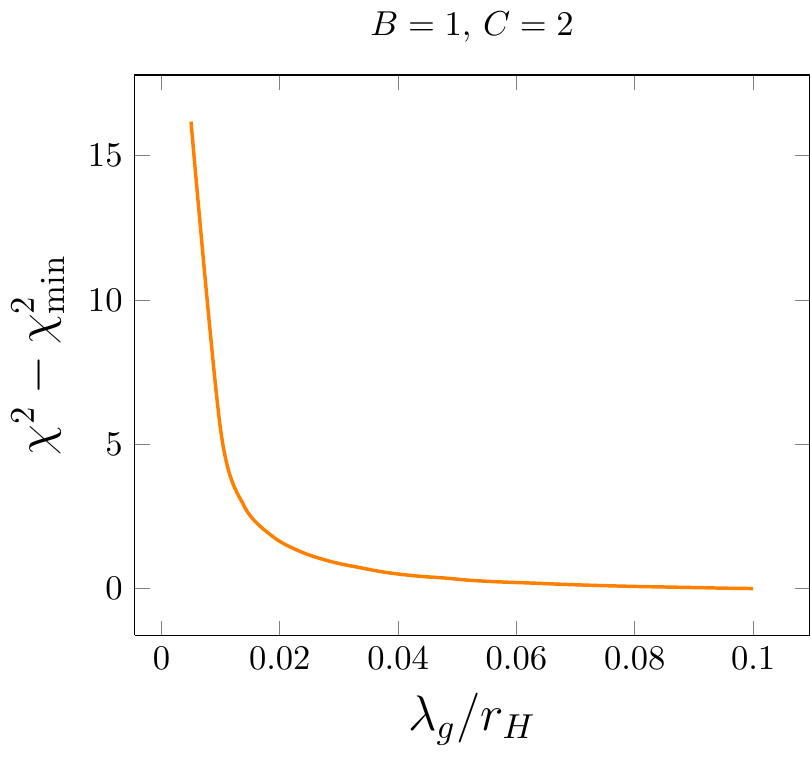}

\caption{\label{fig:stefansln} Cosmological constraints using strong lensing data with 
$B=-1, C=0$ [upper left panel], $B=0, C=1$ [upper right panel],
$B=-2, C=1$ [lower left panel] and $B=1, C=2$ [lower right panel]. The limits
are rather insensitive to the values of $B$ and $C$ and are
on the order of $\lambda_g/r_H\gtrsim 0.01-0.02$ at 95\,\% confidence level.}
\end{center}
\end{figure}

In each fit, we fix the value of $B$ and $C$ and constrain the value of 
$\lambda_g$ in units of the Hubble radius $r_H=c/H_0$.
Results are shown in Figure~\ref{fig:stefansln}. The cosmological limits
are rather insensitive to the values of $B$ and $C$ and are
on the order of $\lambda_g/r_H\gtrsim 0.01-0.02$ at 95\,\% confidence level.

\section{Conclusions} \label{sec:conclusions}

We have analyzed large distance modifications of gravity in the specific context of massive gravity.  
By deriving spherically symmetric solutions, we are able to constrain the parameters of the model using comparison to galaxy observations. 
The theory allows for a Vainshtein mechanism is some parts of the
parameter space. In terms of the normalized couplings of the cubic and
quartic interactions $B=\beta/\alpha^2$ and $C=\gamma / \alpha^3$, the
theory exhibits a Vainshtein mechanism for $C \geq 0$ and $B <
B_\mathrm{max}=\sqrt{5-\sqrt{13}}\sqrt{2C}$. For $C < 0$, general
relativity is not recovered at short distances, and worse for
$B>B_\mathrm{max}$, the solution have pathologies becoming 'multiple
valued' at $B = B_\mathrm{max}$ and 'singular' at $B=\sqrt{3C}$.

Analyzing gravitational lensing data and velocity dispersion measurements of galaxies we put limits on the inverse graviton mass scale $\lambda_g/r_H\gtrsim 0.01-0.02$ at 95\,\% confidence level 
in regions where we do have a Vainshtein mechanism.
This pushes the graviton mass closer to the Hubble radius. An alternative approach would be to use clusters of galaxies acting as gravitational lenses instead of galaxies. In this case, the mass estimate using massive observers would be derived using X-ray data to observe the local pressure gradient inside the cluster to infer the gravitational potential. Although clusters would potentially allow us to probe larger radii than galaxies, we note that such an analysis rely on the assumption of cluster thermal equilibrium. 

\acknowledgments
We would like to thank M. Berg, J. Enander, A. Goobar, J. Gray, F. Hassan, C. de Rham, R. Rosen, A. Schmidt-May, M. von Strauss, B. Sundborg for useful discussions. EM acknowledge support for this study by the Swedish Research Council.

\appendix

\section{Solving for $\pi'$} \label{solutions}

Let us first formulate the problem of solving eq. (\ref{quintic}) using more natural variables. Since we are ultimately interested in the deviations in the force laws, eqs. (\ref{epsilonphi}) and (\ref{epsilonpsi}), let us consider the following dimensionless combination
\be
 \varepsilon \equiv -\frac{\alpha}{\mpl} \frac{\pi'}{\Phi'_\N} \,.
\ee
In the following we will impose the boundary conditions $\varepsilon(r) \to 0$, when $r \to 0$, and $\varepsilon(r) \to 1/3$, when $r \to \infty$, such that we reproduce GR close to the source and FP theory far away from the source. We will also find it convenient to take out the strong coupling scale $r_V = (r_S \lbar^2_g)^{1/3}$, where $r_S = 2 G_\N M$, from the radial variable defining a dimensionless variable\footnote{Not to be confused with the galactic mass density $\rho$.}
\be
 \rho \equiv \frac{r}{r_V} \,.
\ee
Finally, it is more natural to incorporate powers of $\alpha$ to the couplings $\beta$ and $\gamma$, defining
\be
 B \equiv \frac{\beta}{\alpha^2} \,, \quad C \equiv \frac{\gamma}{\alpha^3} \,,
\ee
such that the rescaling $\pi(r) \to \lambda^{-1} \pi(r)$ and $\alpha \to \lambda \alpha$, $\beta \to \lambda^2 \beta$ and $\gamma \to \lambda^3 \gamma$ (which leaves the action in eq. (\ref{action}) invariant) leaves $B$ and $C$ invariant.
 
Using these variables, eq. (\ref{quintic}) for $\pi'$ takes the form
\be \label{simple}
 1 - 3 \varepsilon + 3 \left( \frac{B}{\rho^3} - \frac{C}{4 \rho^6} \right) \varepsilon^2 - \left( \frac{B^2}{2 \rho^6} + \frac{C}{\rho^6} \right) \varepsilon^3 + \frac{3 C^2}{16 \rho^{12}} \varepsilon^5 = 0 \,,
\ee
while the deviations in the potentials, eqs. (\ref{epsilonphi}) and (\ref{epsilonpsi}) take the form
\begin{align}
 \varepsilon_\Phi & = \varepsilon - \frac{C}{4\rho^6} \varepsilon^3 \,, \\
 \varepsilon_\Psi & = \varepsilon - \frac{B}{2\rho^3} \varepsilon^2 + \frac{C}{4\rho^6} \varepsilon^3 \,.
\end{align}
We now consider solutions to eq. (\ref{simple}), first for the two subcases $B \neq 0$, $C=0$ and $B=0$, $C \neq 0$, then attacking the general case $B,C \neq 0$.

\paragraph{Case: ${B \neq 0,\,C=0}$}

For $C=0$, the quintic eq. (\ref{simple}) is reduced to a cubic in $\varepsilon$, which allows for explicit solutions. Absorbing the coupling $B$ into the radial variable, defining $x \equiv |B|^{-1/3} \rho$, eq. (\ref{simple}) is reduced to
\be \label{simpleB}
 1 - 3 \varepsilon + \text{sign}(B) \frac{3 \varepsilon^2}{x^3} - \frac{\varepsilon^3}{2x^6} = 0 \,.
\ee
There are three solutions to this equation, one everywhere real and two somewhere complex solutions. For positive $B$, i.e. $\text{sign}(B)=+1$, the real solution blows up at infinity and we throw this solution away. For negative $B$, i.e. $\text{sign}(B) = -1$ we find a real solution which stays finite at infinity. The real solution for $B<0$ and $C=0$ is given by
\be \label{betasolution}
 \varepsilon(x) = \frac{1}{2} \left[u_B^{1/3}(x) + \frac{8x^6}{u_B^{1/3}(x)} - 4x^3 \right] \,, \quad x = |B|^{-1/3} \rho \,, \quad B<0 \,,
\ee
where $u_B(x) = - 16x^9 + 8x^6 + 4x^3 \sqrt{-16 x^{12} - 16 x^{9} + 4}$. The behavior at small and large $x$ can be inferred directly from the solution, eq. (\ref{betasolution}), or more easily from eq. (\ref{simpleB}). For small $x$ the cubic term dominates while for large $x$ the linear term does, and we find
\be
 \lim_{x \to 0} \varepsilon(x) = 2^{1/3} x^2 \,, \quad \lim_{x \to \infty} \varepsilon = \frac{1}{3} \,.
\ee
Thus in the limit of small $x$ we reproduce GR while in the limit of large $x$ we reproduce FP theory.

\paragraph{Case: ${B=0,\,C \neq 0}$}

In the case $B = 0$, eq. (\ref{simple}) takes the form
\be \label{simpleC}
 1 - 3 \varepsilon - \text{sign}(C) \frac{3 \varepsilon^2}{4 y^6} - \text{sign}(C) \frac{\varepsilon^3}{y^6} + \frac{3 \varepsilon^5}{16 y^{12}} = 0 \,,
\ee
where $y \equiv |C|^{-1/6} \rho$. We now argue that in this case, the quintic term is of no importance. First, we demand that at infinity, $\varepsilon$ goes to $1/3$. Then clearly $\varepsilon^5/y^{12}$ is highly suppressed compared to the other terms for large $y$. Second, for small $y$ we demand that $\varepsilon$ goes to 0. If the quintic should give any contribution at all, then $\varepsilon \, \propto \, y^{12/5}$ at the origin, in which case the quadratic $\varepsilon^2/y^6 \, \propto \, y^{24/5-6}$ blows up and dominates the behavior at the origin. Thus the quintic term is of no importance neither at the origin or at infinity. We then solve the cubic equation, neglecting the quintic term, and simply observe that on the solution the quintic is negligible for all $y$, with maximal corrections $\mathcal{O}(10^{-6})$ around $y=1$. The cubic equation has one everywhere real solution and two somewhere complex solutions. For negative $C$, i.e. $\text{sign}(C) = -1$, the solution does not go to zero at the origin, and we throw this solution away. For positive $C$, i.e. $\text{sign}(C) = +1$ we find a well-behaved solution 
\be \label{gammasolution}
 \varepsilon(y) = \frac{1}{4} \left[ u_C^{1/3}(y) + \frac{1-16 y^6}{u_C^{1/3}(y)} -1 \right] \,, \quad y = |C|^{-1/6} \rho \,, \quad C>0 \,,
\ee
where $u_C(y) = 56 y^6 - 1 + 8 y^3 \sqrt{64 y^{12} + 37 y^6 - 1}$. It is now easy to see, either directly from the solution, eq. (\ref{gammasolution}), or from eq. (\ref{simpleC}), that we indeed have
\be
 \lim_{y \to 0} \varepsilon(y) = \frac{2}{\sqrt{3}} y^3 \,, \quad \lim_{y \to \infty} \varepsilon(y) = \frac{1}{3} \,.
\ee

\paragraph{Case: ${B,C \neq 0}$}
In the case of non-zero $B$ and $C$ the situation is more subtle. Let us first analyze the parameter range of $C$. Now, observe that for $C \neq 0$, the small $\rho$ behavior is dominated by the terms $1$ and $-\frac{3C}{4 \rho^6} \varepsilon^2$ in eq. (\ref{simple}). Thus, we conclude that $\varepsilon \to \frac{2}{\sqrt{3C}} \rho^3$ when $\rho \to 0$, which implies $C>0$. We can than analyze the parameter range of $B$ given some $C>0$. We split up the case in $B<0$ and $B>0$.

\paragraph{\emph{Subcase}: ${C>0,\,B<0}$}
For $C>0$ and $B<0$ the situation is very much the same as in the already considered above cases. We find a solution to eq. (\ref{simple}) (neglecting the quintic term) with the right behavior at the origin and at infinity which simply interpolates between eqs. (\ref{betasolution}) and (\ref{gammasolution})
\be \label{generalsolution}
 \varepsilon(\rho) = \frac{1}{2(2C+B^2)} \left[ u^{1/3}(\rho) + \frac{w(\rho)}{u^{1/3}(\rho)} + v(\rho) \right] \,,
\ee
where we have defined
\begin{align}
 u(\rho) & = g(\rho) + 4 (2C + B^2) \rho^3 \sqrt{f(\rho)} \,, \\
 v(\rho) & = 4 B \rho^3 - C \,, \\
 w(\rho) & = (-16C + 8 B^2) \rho^6 - 8 B C \rho^3 + C^2 \,, \\
 f(\rho) & = 16(4 C - B^2) \rho^{12} + 8 (2 B^3 - 9 BC) \rho^9 \\ & + (4 B^4 + 37 C^2 - 20 B^2 C) \rho^6 + 12 B C^2 \rho^3 - C^3 \,, \\
 g(\rho) & =  16(B^3 - 6 BC ) \rho^9 + 4(2 B^4 - B^2 C + 14 C^2) \rho^6 + 12 B C^2 \rho^3 -C^3 \,.
\end{align}
Just as the case $C \neq 0$, $B=0$ we find that the quintic term contributes negligible on the solution and eq. (\ref{generalsolution}) is indeed a good approximation to the full solution.

\paragraph{\emph{Subcase}: ${C>0,\,B>0}$} We then try solving eq. (\ref{simple}) for $B>0$, neglecting the quintic term. This time, the solution to the cubic equation is only a good approximation for $B \lesssim B_\mathrm{max}$, where $B_\mathrm{max} = \sqrt{5-\sqrt{13}}\sqrt{2C}$ as we show below. For $B \lesssim B_\mathrm{max}$ we find a solution with the correct limiting behavior
\be \label{generalsolution2}
 \epsilon(\rho) =
 \left\{
 \begin{array}{ll}
 \frac{1}{2(2C+B^2)} \left[ u^{1/3}(\rho) + \frac{w(\rho)}{u^{1/3}(\rho)} + v(\rho) \right] & \text{for } \rho \leq \rho_\star \\
 \frac{1}{2(2C+B^2)} \left[ e^{i\pi/3} u^{1/3}(\rho) - e^{-i\pi/3} \frac{w(\rho)}{u^{1/3}(\rho)} + \frac{v(\rho)}{2} \right] & \text{for } \rho > \rho_\star
 \end{array}
 \right. \,,
\ee
where $\rho_\star$ is the zero of $f(\rho_\star) = 0$. We now show that for $B \simeq B_\mathrm{max}$ the quintic term induces important corrections.

\paragraph{Geometric formulation:}

\begin{figure}
\begin{center}
\includegraphics[angle=0,width=.5\textwidth]{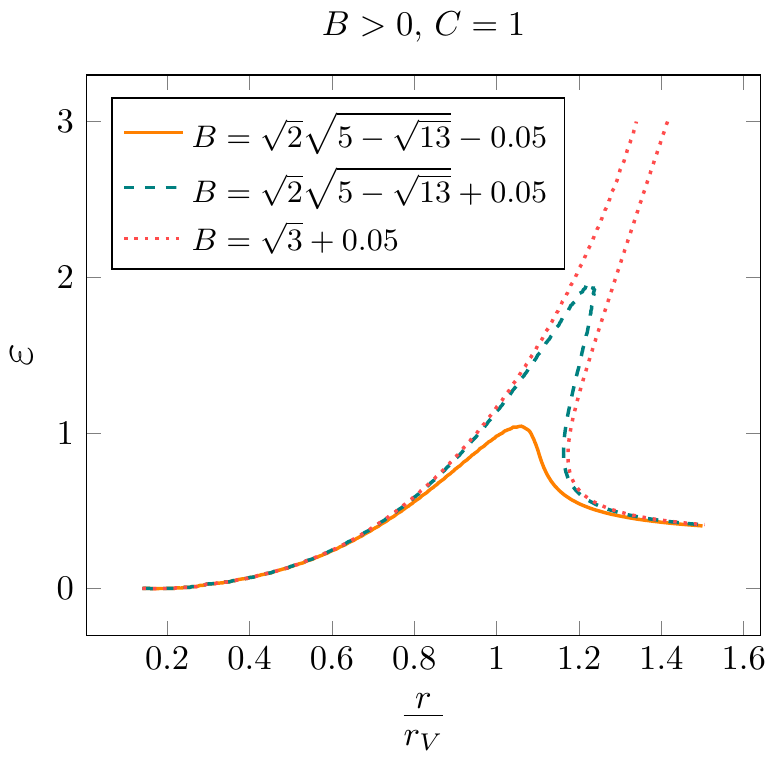}
\caption{As we vary the model parameter $B \equiv \frac{\beta}{\alpha^2}$, while keeping $C \equiv \frac{\gamma}{\alpha^3}$ fixed, the curve $\varepsilon = \varepsilon(r/r_V)$ develops a singular behavior at $B = \sqrt{3C}$ where the curve 'opens up', see pink dotted line above. Already before this happens the curve becomes 'multiple valued' at $B =\sqrt{5-\sqrt{13}}\sqrt{2C}$, see the dashed curve which acquires three values for $\varepsilon(r/r_V)$ in the region $r/r_V \simeq 1.2$. For $B<\sqrt{5-\sqrt{13}}\sqrt{2C}$ the curve is well-behaved, see orange solid curve. The above example highlights Vainshtein's original idea: Models with a well-behaved short and long-distances behavior might or might not analytically connect the two regions.}
\label{fig:bound}
\end{center}
\end{figure}

Let us consider the problem in more geometric terms. The curve $\varepsilon = \varepsilon(\rho)$ in the $\rho\varepsilon$-plane is the zero locus of the following polynomial in $\rho,\varepsilon$ with moduli $B,C$
\be
 p(\varepsilon,\rho; B,C) = 1 - 3 \varepsilon + 3 \left( \frac{B}{\rho^3} - \frac{C}{4 \rho^6} \right) \varepsilon^2 - \left( \frac{B^2}{2 \rho^6} + \frac{C}{\rho^6} \right) \varepsilon^3 + \frac{3 C^2}{16 \rho^{12}} \varepsilon^5 \,.
\ee
As we vary the moduli $B,C$ the curve is changing shape and we might encounter singularities, i.e. points where the tangent vanishes, see Figure \ref{fig:bound} for reference. To find potential singularities we analyze the system of equations
\be \label{singular}
 p = 0\,, \quad \p_\rho p =0 \,, \quad \p_\varepsilon p = 0 \,.
\ee
Using standard methods we can find a Groebner basis where the variables $\rho,\varepsilon$ are eliminated, and the system in eq. (\ref{singular}) is reduced to
\be
 B^4 - 15 B^2 C + 36 C^2 = 0 \,.
\ee
This equation has solutions $B^2 = 3C$ and $B^2 = 12C$, and the solution we are seeking is given by $B = \sqrt{3C} \approx 1.73 \sqrt{C}$. Thus, we find that the curve $\varepsilon=\varepsilon(\rho)$ encounters a singularity at $B = \sqrt{3C}$ and eq. (\ref{simple}) definitely has no solutions for $B>\sqrt{3C}$. In fact, even before reaching this singularity there is a point where the curve becomes multiple-valued. At this point the tangent to the curve $\p_\rho \varepsilon$ goes to infinity. Thus by considering the system
\be
 p = 0 \,, \quad \left(\p_\rho \varepsilon \right)^{-1} = 0 \,, \quad \left(\p^2_\rho \varepsilon \right)^{-1} = 0\,,
\ee
we find using similar methods as above that
\be
 4 B^{12} - 72 B^{10} C - 56 B^8 C^2 + 3220 B^6 C^3 - 28627 B^4 C^4 + 109308 B^2 C^5 - 138384 C^6 = 0\,.
\ee
The solution of interest for us is $B = \sqrt{5-\sqrt{13}}\sqrt{2C} \approx 1.67 \sqrt{C}$, thus there are no solutions to eq. (\ref{simple}) for $B> B_\mathrm{max} \equiv \sqrt{5-\sqrt{13}}\sqrt{2C}$. This situation is depicted in Figure \ref{fig:bound}.

\paragraph{Summary:}

We end the appendix by summarizing our findings. To find solutions with the right boundary conditions then $C \geq 0$. For $B \leq 0$ and $0 < B \lesssim B_\mathrm{max}$, where $B_\mathrm{max} =\sqrt{5-\sqrt{13}}\sqrt{2C}$, the solutions to the quintic equation are well approximated by the solutions to the cubic equation with at most $\mathcal{O}(10^{-6})$ corrections. At $B \simeq B_\mathrm{max}$ the solution to the cubic approximates the solution to the quintic with corrections as big as up to $40\%$. Beyond $B>B_\mathrm{max}$ there are no solutions to the quintic equation satisfying the specified boundary conditions.

\bibliographystyle{unsrt}
\bibliography{refs.bib}

\end{document}